\documentclass[preprint,showpacs,preprintnumbers,amsmath,amssymb]{revtex4}

\usepackage{graphicx}
\usepackage{dcolumn}
\usepackage{bm}

\newcommand{\Glam}{$\Gamma_{\Lambda}$}
\newcommand{\Gtot}{$\Gamma_{tot}$}
\newcommand{\Gm}{$\Gamma_{m}$}
\newcommand{\Gpim}{$\Gamma_{\pi^{-}}$}
\newcommand{\Gpin}{$\Gamma_{\pi^{0}}$}
\newcommand{\Gmr}{$\Gamma_{\pi^{0}} / \Gamma_{\pi^{-}}$}
\newcommand{\Gnm}{$\Gamma_{nm}$}
\newcommand{\Gp}{$\Gamma_{p}$}
\newcommand{\Gn}{$\Gamma_{n}$}
\newcommand{\GnGp}{$\Gamma_{n} / \Gamma_{p}$}
\newcommand{\kpi}{$(K^{-}, \pi^{-})$}
\newcommand{\pik}{$(\pi^{+}, K^{+})$}

\newcommand{\pipp}{$(\pi^{+},pp')$}
\newcommand{\pippi}{$(\pi^{+}, p\pi)$}
\newcommand{\Cpik}{$^{12}$C$(\pi^{+},K^{+})$}
\newcommand{\Sipik}{$^{28}$Si$(\pi^{+},K^{+})$}
\newcommand{\Fepik}{Fe$(\pi^{+},K^{+})$}
\newcommand{\slam}{s$_{\Lambda}$}
\newcommand{\plam}{p$_{\Lambda}$}

\newcommand{\Hlam}{$^{4}_{\Lambda}$H}
\newcommand{\Helam}{$^{4}_{\Lambda}$He}
\newcommand{\Hegolam}{$^{5}_{\Lambda}$He}
\newcommand{\Clam}{$^{12}_{\Lambda}$C}
\newcommand{\Blam}{$^{11}_{\Lambda}$B}
\newcommand{\Silam}{$^{28}_{\Lambda}$Si}
\newcommand{\Allam}{$^{27}_{\Lambda}$Al}
\newcommand{\Felampre}{$^{56}_{\Lambda}$Fe}
\newcommand{\Felam}{$_{\Lambda}$Fe}
\newcommand{\Ylam}{$^{89}_{\Lambda}$Y}

\begin{document}


\title{ Mesonic and nonmesonic weak decay widths of medium-heavy $\Lambda$ 
hypernuclei }

\author{ Y.~Sato,$^{1,3}$ S.~Ajimura,$^{5}$ K.~Aoki,$^{3}$ H.~Bhang,$^{2}$ 
T.~Hasegawa,$^{4,a}$, O.~Hashimoto,$^1$, H.~Hotchi,$^{4,b}$\\ 
Y.D. Kim,$^{2,3,c}$  T.~Kishimoto,$^{5}$ K.~Maeda,$^{1}$
H.~Noumi,$^{3}$ Y.~Ohta,$^{4}$ K. Omata,$^{3}$ H. Outa,$^{3}$\\ 
H. Park,$^{2}$
M. Sekimoto,$^{3}$ T. Shibata,$^{3}$ T. Takahashi,$^{1,3}$ and M.~Youn$^{2,3}$.\\
}
\address{
$^{1}$Department of Physics, Tohoku University, Sendai 980-8578, Japan\\
$^{2}$Department of Physics, Seoul National University, Seoul 151-742, Korea\\
$^{3}$High Energy Accelerator Research Organization (KEK),
Tsukuba, Ibaraki 305-0801, Japan\\
$^{4}$Graduate School of Science, University of Tokyo, Tokyo
113-0033, Japan\\
$^{5}$Department of Physics, Osaka University, Toyonaka, Osaka 560-0043, Japan \\
$^{a}$School of Allied Health Sciences, Kitazato University,
Sagamihara 228-8555, Japan.\\
$^{b}$Research Center for Physics and Mathematics,
Osaka electro-communication University, Neyagawa, Osaka 572-8530, Japan \\
$^{c}$Department of Physics, Sejong University, Seoul 143-747, Korea
}  
\date{\today}

\begin{abstract}

We have measured the energy spectra of pions and protons emitted in the weak decay of 
\Clam, \Silam, and \Felam~hypernuclei produced via the \pik~reaction. 
The decay widths of the $\pi^{-}$ mesonic decay $(\Lambda \rightarrow p \pi^{-})$
and the nonmesonic decay $(\Lambda N \rightarrow N N)$ were extracted.
The present results demonstrate an increase of the mesonic decay width due to
a distortion of the pion wave function in nuclear medium for the first time. 
The ratios of the neutron- to proton-induced nonmesonic decay widths, 
$\Gamma_{n}(\Lambda n \rightarrow n n)/\Gamma_{p}(\Lambda p \rightarrow n p)$,
were evaluated by a direct comparison of the measured proton energy spectra
with the calculated ones.
No theoretical calculation which has been proposed so far can simultaneously account for
both the nonmesonic decay widths and the \GnGp~ratios in the present data.

\end{abstract}


\maketitle

\narrowtext
\section{ Introduction }
\label{sec:intro}

The $\Lambda$ hypernuclei have been extensively studied since their first discovery
in a nuclear emulsion~\cite{Dan53}.
A $\Lambda$ hyperon, which carries a degree of freedom, `strangeness', can be
used as a unique probe to investigate the interior of the nucleus
since it does not suffer from Pauli's exclusion principle.

A $\Lambda$ hypernucleus is usually produced in an excited state
of $\Lambda$-particle neutron-hole configuration.
When a $\Lambda$ hypernucleus is excited above the particle emission
threshold, it decays dominantly by the strong interaction, and then deexcites
to its ground state via electromagnetic transitions. 
From the ground state, it eventually decays through the weak interaction.

The total decay width (\Gtot = 1/$\tau_{HY}$) of a $\Lambda$ hypernucleus
is composed of the mesonic decay width (\Gm) and the nonmesonic
decay width (\Gnm).
In the mesonic decay, a $\Lambda$ hyperon decays into a nucleon and
a pion in nuclear medium, similarly as in free space.
The mesonic decay width (\Gm) can be further expressed as a sum of the decay widths
for emitting negative (\Gpim: $\Lambda \rightarrow p\pi^{-}$) and neutral 
(\Gpin: $\Lambda \rightarrow n\pi^{0}$) pions, respectively.
In two-body nonmesonic decay, however, a $\Lambda$ hyperon in a nucleus
interacts with a neighboring nucleon and decays into a pair of nucleons
without emitting a pion.
The nonmesonic decay width (\Gnm) is comprised of the proton-induced
(\Gp: $\Lambda p \rightarrow np$) and neutron-induced
(\Gn: $\Lambda n \rightarrow nn$) decay widths.
In addition, importance of the two nucleon-induced decay width
($\Gamma_{2N}$: $\Lambda N N \rightarrow N N N$) has also been discussed
theoretically~\cite{Eric92}, although it has not been experimentally established yet.

Therefore, the lifetime ($\tau_{HY}$), the total (\Gtot) and partial 
decay widths of $\Lambda$ hypernuclei are connected by the following relationship,
\begin{eqnarray}
  1 / \tau_{HY} = 
  \Gamma_{tot} &=& \Gamma_{m} + \Gamma_{nm}, \\
    \Gamma_{m} &=& \Gamma_{\pi^{-}} + \Gamma_{\pi^{0}}, \\
   \Gamma_{nm} &=& \Gamma_{p} + \Gamma_{n} (+ \Gamma_{2N}).
\end{eqnarray}

The mesonic decay releases energy of $Q^{free} \simeq 40$ MeV,
corresponding to a momentum of about 100 MeV/{\it c} in the center-of-mass system
(c.m.) for the emitted nucleon and pion.
Since this is much smaller than the nuclear Fermi momentum
($k_{F} \sim 270$ MeV/{\it c}) in typical nuclei, except for light ones, 
the mesonic decay is strongly suppressed in heavier hypernuclei due to
Pauli's exclusion principle acting on the nucleon in the final state.

Although there are several experimental data of \Gpim~and/or \Gpin~on light $\Lambda$
hypernuclei, such as \Hlam~\cite{Outa95}, \Helam~\cite{Outa98}, 
\Hegolam~\cite{Szym91}, \Blam, and
\Clam~\cite{Szym91}\cite{Saka91}\cite{Noumi95}, no finite value has
been reported for the mesonic decay widths of $\Lambda$ hypernuclei heavier
than \Clam.
It has been pointed out that the mesonic decay widths are sensitive to
the nuclear structure in the final state and 
the choice of the pion-nucleus optical potentials~\cite{Motoba88_NPA}\cite{Motoba94}\cite{Nieves93}.
While the pion-nucleus potentials have been studied so far through
$\pi$-nucleus scattering experiments and measurements of X-rays from the pionic
atoms, the study of mesonic weak decay offers a unique opportunity to
investigate pion wave functions deep inside a nucleus, where $\Lambda$ hypernuclear
weak decay occurs.
Therefore, precise and systematic measurements of the mesonic decay widths
are long awaited.

Nonmesonic weak decay provides a unique opportunity to study
both the parity-conserving and parity-violating terms in the
baryon-baryon weak interaction, which are difficult to experimentally
investigate by other means, since the weak component of nucleon-nucleon scattering
is usually masked by the strong interaction.

In nonmesonic decay of a $\Lambda$ hypernucleus, the released energy is
$Q^{free} \simeq 176$ MeV, and the c.m. momentum of two nucleons 
in the final state is about 400 MeV/{\it c}.
Since the momentum transfer is larger than the Fermi momentum, 
the nonmesonic decay does not suffer from Pauli blocking 
so seriously.
Therefore, it dominates over the mesonic decay, except in very light
$\Lambda$ hypernuclei.

Currently, there is no clear picture concerning the nonmesonic decay,
although various experimental and theoretical efforts have been attempted
for a long time.
Nonmesonic decay width (\Gnm), the neutron- and proton-induced partial 
decay widths (\Gn~and \Gp), and their ratio (\GnGp) are thought to be good 
experimental observables.


The lifetimes and partial decay widths of several spin-isospin saturated 
$\Lambda$ hypernuclei were measured at Brookhaven National
Laboratory (BNL) and High Energy Accelerator Research Organization (KEK)
(\Hegolam~\cite{Szym91}, \Blam, and \Clam~\cite{Szym91}\cite{Noumi95}).
It was reported that the \GnGp~ratios were close to unity or larger, 
although the quoted errors were large.
Contrary to the experimental data, theoretical calculations based
on the framework of meson exchange models assuming the $|\Delta I| = 1/2$ rule 
yield \GnGp~ratios much smaller than unity (typically 0.1-0.3)
\cite{Dubach86}\cite{Bando90}\cite{Parreno_PRC97}.
The smallness of the \GnGp~ratios essentially comes from
the tensor dominance of one pion-exchange,
which contributes only to the proton-induced nonmesonic decay. 
The inclusion of the $2\pi/\rho$ exchange processes was discussed in an attempt
to cancel out the strong tensor dominance of one pion-exchange
potential~\cite{Ito98}.
Another approach based on the direct quark current exchange has also 
been suggested to take the short-range nature of the nonmesonic decay
and the violation of the $|\Delta I| = 1/2$ rule into account~\cite{Oka}.
All of these efforts still resulted in small \GnGp~ratios.
Good experimental data are seriously required in order to reveal
the mechanism of the nonmesonic decay.

We carried out an experiment to measure the precise lifetimes and partial decay
branching ratios of \Clam, \Silam~and \Felam~hypernuclei produced via
the \pik~reaction (KEK-PS E307).
Since the contribution of mesonic decay to the total decay width becomes
smaller, measurements of the nonmesonic decay widths becomes less ambiguous
in heavier $\Lambda$ hypernuclei.
The results of lifetime measurements in the present experiment
have already been published elsewhere~\cite{Bhang98}\cite{Park99}.
The present paper reports results for the $\pi^{-}$ mesonic decay widths
(\Gpim), the total nonmesonic decay widths (\Gnm), and the \GnGp~ratios of
\Clam, \Silam, and \Felam~hypernuclei, a part of which has already been
published in ref.~\cite{E307pdw_letter}.

The experimental setup is described in Sec.~\ref{sec:exp}. 
The data analysis and results are explained in Sec.~\ref{sec:result} and
discussed along with previous studies in Sec.~\ref{sec:discuss}.
The conclusion is described in Sec.~\ref{sec:concl}.

\section{ Experiment }
\label{sec:exp}

The $\Lambda$ hypernuclei produced by the \pik~reaction were identified
in the hypernuclear mass spectrum reconstructed using the momenta of
the incident pions and the outgoing kaons.
The charged particles emitted in the weak decay were detected by the decay
counter systems located above and below the experimental target
in coincidence with kaons.
The branching ratios of the $\pi^{-}$ mesonic decay were obtained as ratios
between the numbers of identified $\Lambda$ hypernuclei and pions emitted 
after the mesonic decay.
The $\pi^{-}$ mesonic decay width (\Gpim) and the nonmesonic decay width (\Gnm) 
were then derived by combining the results of the present lifetime measurement
and the previous data of the $\pi^{0}$ mesonic decay branching ratios.
The \GnGp~ratios were evaluated by a direct comparison of the observed 
proton energy spectra with calculated ones as described in the next section.

The \pik~reaction has several advantages in the measurement of the 
weak decay of heavy $\Lambda$ hypernuclei.
On the other hand, the \kpi~reaction, whose momentum transfer is very small,
dominantly excited substitutional states and poorly populates the bound region of
a $\Lambda$ hypernucleus~\cite{Povh80}.
On the contrary, the \pik~reaction, whose momentum transfer is about 
400 MeV/{\it c}, is suitable to excite deeply bounded states of a
$\Lambda$ hyperon.
Another advantage of the \pik~reaction over the \kpi~reaction is that 
the pion beams are relatively clean compared with kaon beams which is
subject to a large contamination of pions originated in the in-flight
decay.
It is crucial for a precise measurement of the branching ratios for heavier $\Lambda$
hypernuclei such as \Silam~to obtain inclusive spectra in which each peak 
is clearly separated with high statistics and less background contamination.

\begin{figure}[htb]
\caption{ Schematic view of the KEK-PS K6 beam line and SKS. }
\includegraphics[width=10.0cm]{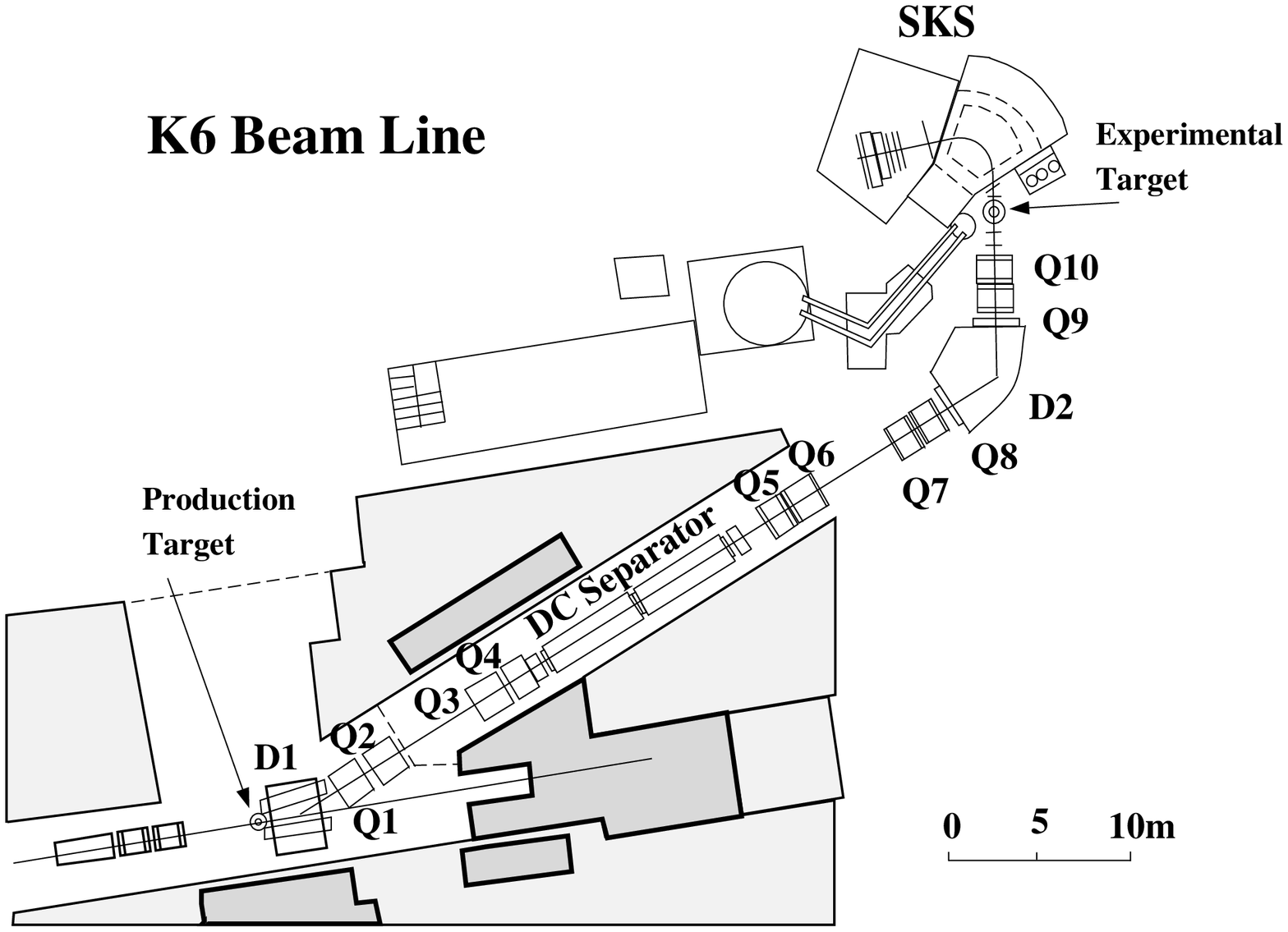}
\label{k6bl}
\end{figure}

The present experiment was carried out at the K6 beam line~\cite{Tanaka_K6}
of the KEK 12-GeV Proton Synchrotron (PS), which is depicted in Fig.~\ref{k6bl}.
Primary proton beams were extracted for 1.4-1.8 sec in each 4.0 sec period.
The central momentum of pion beams was set to 1.06 GeV/{\it c}, where
the production cross section of the elementary \pik~reaction becomes maximum~\cite{Dover80}.
The typical beam intensity at the target was adjusted to $3.6 \times 10^{6}$ 
$\pi^{+}$ per 1.8 sec for $2.0 \times 10^{12}$ primary protons in order to 
maintain the stability of the total detection efficiency of the tracking
chambers on the beam line.
The accumulated numbers of incident pions used in the present analysis are listed 
in Table~\ref{yieldtab}.

Plates of graphite ($^{nat}$C), natural silicon ($^{nat}$Si) and
iron ($^{nat}$Fe) were used as the experimental targets.
These targets were tilted up by $10-15$ degrees from the beam direction
in order to maximize the thickness in the beam direction ($6-10$~$ g/cm^{2}$)
and minimize the threshold energy of the charged decay particles.
Detailed specifications of the experimental targets can be found in ref.~\cite{Park99}.

The momenta of the scattered particles were analyzed by the Superconducting Kaon 
Spectrometer (SKS).
SKS has specifications particularly suitable for coincidence experiments
of hypernuclei, such as a large solid angle (100 {\it msr}) and short flight
length ($\sim 5$ {\it m}), as well as a good momentum resolution
(FWHM $\sim$ 0.1\% at 720 MeV/{\it c})~\cite{SKSNIM}.
Kaon events were selected in the mass spectrum of scattered particles,
which were reconstructed by their time-of-flight and momentum.
The mass of $\Lambda$ hypernuclei was calculated for selected \pik~events through
the momentum vectors of incident pions and scattered kaons with corrections of
the energy loss in the beam counters and the experimental target.
The correction of the horizontal scattering angle measured by SKS was applied
to the momenta of outgoing kaons.
In addition, the constant offsets were applied to the present hypernuclear mass spectra
so as to adjust the measured ground states to the ones measured by the previous
studies~\cite{Davis92}\cite{Hase96}\cite{Akei91}.
The constant offsets applied for the spectra with the \Cpik, \Sipik, and \Fepik reactions
were 0.62 MeV, 3.51 MeV, and 2.0 MeV, respectively.

\begin{figure}[htb]
\caption{ Schematic view of the decay counter system. }
\vspace*{0.1in}
\includegraphics[width=10.0cm]{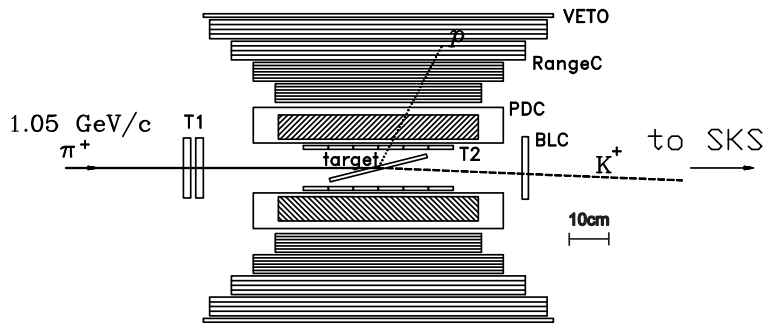}
\label{decayfig}
\end{figure}

Charged particles emitted in the weak decay of $\Lambda$ hypernuclei 
were detected by the decay counter systems located above and below 
the experimental target, as shown in Fig.~\ref{decayfig}.
Each system was comprised of timing scintillation counters (T1 and T2), 
multi-wire drift chambers (PDCU and D), and range counters (RangeU and D).
The plastic scintillation counters (T1 and T2) were designed to realize
good time resolution ($\sigma \sim 40$ psec) even under
the high-rate of $10^{6}$ particle/sec~\cite{YDKim}.
The T1 counters installed on the beam line gave the time of hypernuclear production,
and the T2 counters located above and below the target 
measured time information and deposited energy $(\Delta E)$ of the decay particles.
A detailed description of the fast timing measurement in the present 
experiment can be found in ref.~\cite{Park99}.

The PDC is a multi-wire drift chamber having two plane-pairs parallel 
and one plane-pair perpendicular to the beam direction.
The distance between the sensitive wires in each plane was 8 {\it mm}, and the
elements of each plane-pair were displaced relative to the other by a half of
the wire spacing to solve the left/right ambiguity.
A typical spatial resolution of PDC was $ \sigma \simeq 300 \mu$m.
Tracks of the decay particles were reconstructed by fitting the hit positions on PDC
and the reaction vertex on the target plane.

The range counter is comprised of twenty plastic scintillators, 
of which 12 slabs had 4 {\it mm} thickness and 8 slabs 6 {\it mm} 
thickness, respectively.
The region of kinetic energy measured by the range counter system was
from 30 MeV to 150 MeV for protons, and from 12 MeV to 70 MeV for pions, respectively.
The total kinetic energy deposited in the range counters ($E_{tot}$) was also
evaluated by summing up all ADC data after adjusting pulse height gain of each segment.

\begin{figure}[htb]
\caption{ The particle identification functions in the region of the \Clam~ground 
state with fitting curves. 
  (a) PID1 function made from $\Delta E$ and range.
  (b) PID2 function made from $\Delta E$ and E$_{tot}$.}
\vspace*{0.1in}
\includegraphics[width=7.5cm]{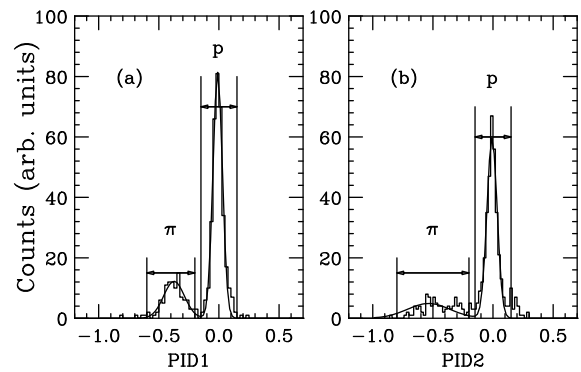}
\label{pidfit_c}
\end{figure}

Pions and protons were identified with the functions (PID1 and PID2) made from
the measured $\Delta E$, range, and $E_{tot}$.
These functions were defined as follows:
\begin{eqnarray}
  X &=& \ln(range),  \nonumber \\
  Y &=& \ln (dE / dx), \nonumber \\
  Z &=& \ln(E_{tot}), \nonumber \\
  PID1 &=& Y - P_{2}(X), \\
  PID2 &=& Y - P_{2}(Z).
\end{eqnarray}
Here, the mean energy loss ($dE / dx$) of each decay particle was calculated
by the energy deposit $(dE)$ with a correction of the flight length $(dx)$ in the T2 counters.
The symbol $P_{2}$ denotes the second-order polynomial function, which was used
to correct the correlations and make the one-dimensional plot of
particle identification functions (PID functions).
Protons and pions which passed the criteria of both PID1 and PID2 functions were
used in the analysis.
Fig.~\ref{pidfit_c} shows the PID functions obtained in the \Clam~ground state region
with Gaussian-shape fitting curves.
The accepted efficiency of the PID window for pions and protons is $(98.8 \pm 1.2)$\%, and
it was included in the estimation of detection efficiency.
The proton events in the pion window were evaluated to be less than 1\% of pion
events, after applying both PID1 and PID2 gates.

The overall solid angle of the two decay counter systems was estimated 
to be $\Omega_{coin}/ 4\pi  = (27 \pm 1)$~\% for \Clam~and \Felam, and
$\Omega_{coin}/4 \pi = (28 \pm 1)$~\% for \Silam~by a Monte-Carlo simulation
based on GEANT~\cite{GEANT}.
The error of the solid angle comes from the angle resolution of the drift chambers.
The total detection efficiency ($\varepsilon_{coin}$) of each decay counter system,
including the ambiguity of the particle identification, was estimated to be 
($84 \pm 2$)~\% for the carbon and iron targets, and ($86 \pm 2$)~\% for the silicon 
target.
These efficiencies were obtained by analyzing the calibration data of the \pipp~and
\pippi~reactions, which were taken simultaneously with the \pik~trigger data.

The stability of the detection efficiency during the experiment was
studied by monitoring the detection efficiency in each data-taking run, which
usually took about 2 hour.
The errors in the detection efficiency are dominantly due to statistics
of the calibration data.
These errors in the evaluation of the  solid angle and the detection 
efficiency are included in the systematic errors of the branching ratios.

\section{ Analysis and Results }
\label{sec:result}

\subsection{ Hypernuclear mass spectra }

\begin{figure}
\caption{ Hypernuclear mass spectra by the \Cpik~reaction. 
  (a) inclusive,
  (b) with coincident protons ($E_{p} > 40$ MeV), and 
  (c) with coincident pions ($E_{\pi} > 12.5$ MeV). }
\includegraphics[width=7.5cm]{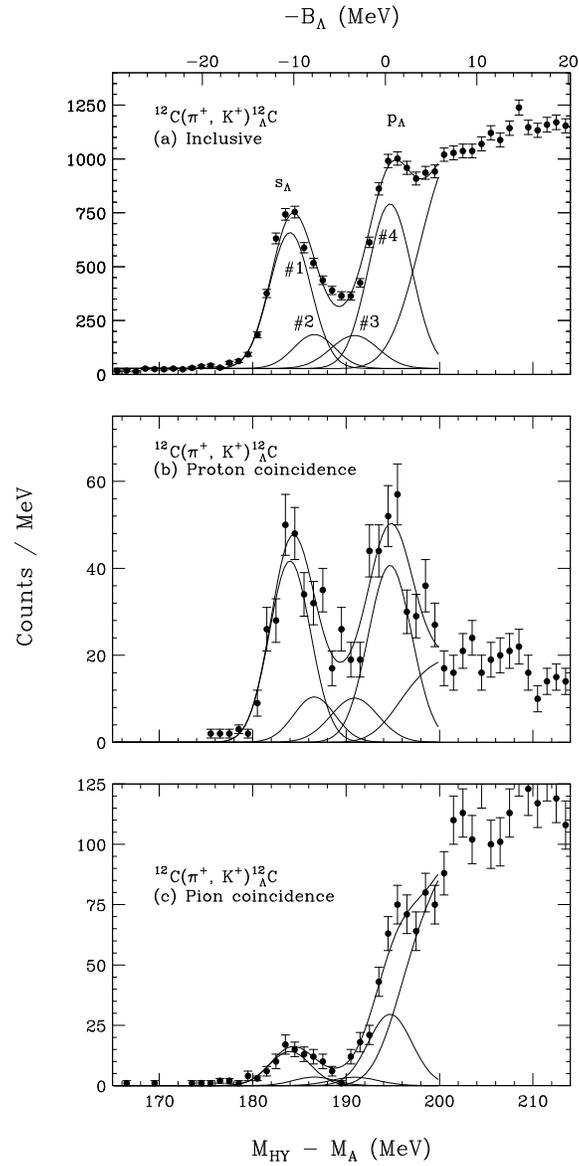}
\label{massfig_c}
\end{figure}

\begin{figure}
\caption{ Hypernuclear mass spectra by the \Sipik~reaction. 
  (a) inclusive,
  (b) with coincident protons ($E_{p} > 40$ MeV), and 
  (c) with coincident pions ($E_{\pi} > 12.5$ MeV). }
\includegraphics[width=7.5cm]{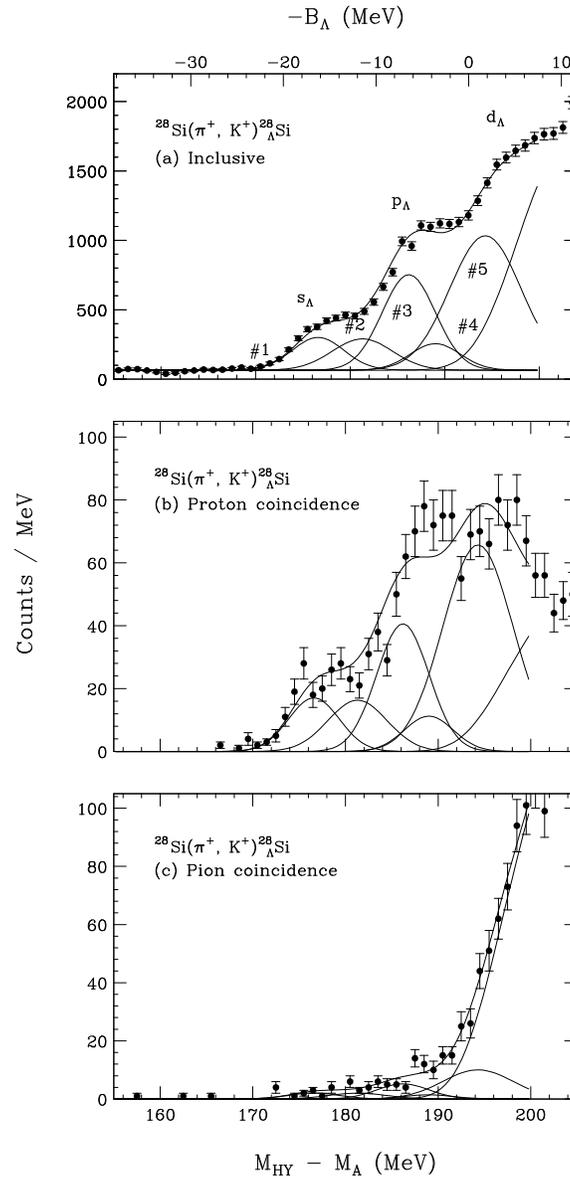}
\label{massfig_si}
\end{figure}

\begin{figure}
\caption{ Hypernuclear mass spectra by the \Fepik~reaction. 
  (a) inclusive,
  (b) with coincident protons ($E_{p} > 40$ MeV), and 
  (c) with coincident pions ($E_{\pi} > 12.5$ MeV). }
\includegraphics[width=7.5cm]{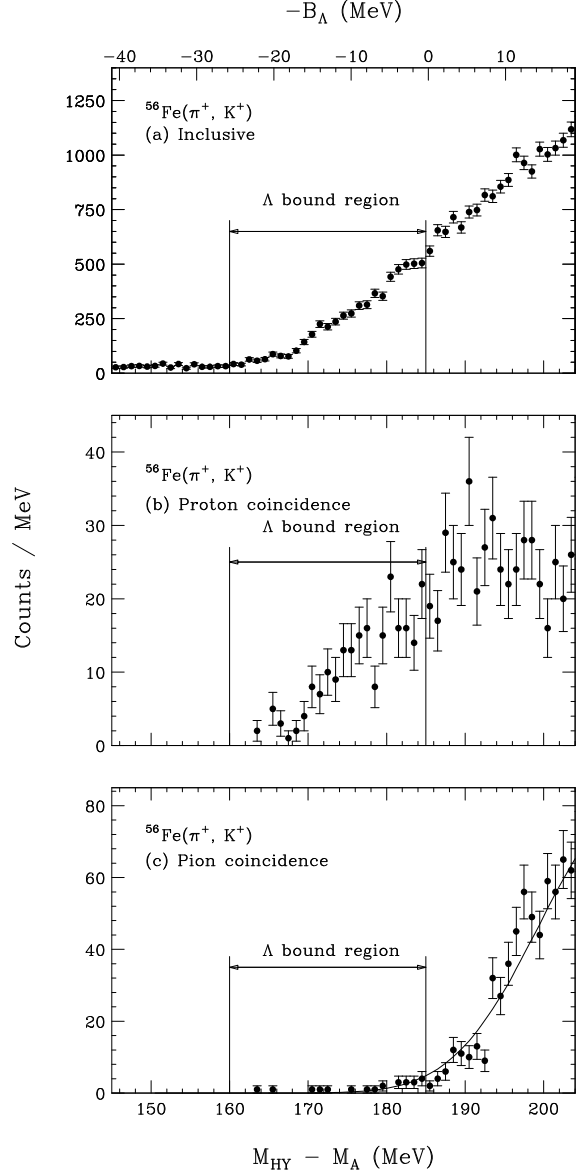}
\label{massfig_fe}
\end{figure}

Fig.~\ref{massfig_c},~\ref{massfig_si}, and \ref{massfig_fe} illustrate 
the measured hypernuclear mass spectra.
The horizontal axis is the mass difference between a produced $\Lambda$ hypernucleus
and a target nucleus ($M_{HY} - M_{A}$).
A scale in the binding energy of the $\Lambda$ hyperon (B$_{\Lambda}$)
is also given in the same figure.
The top spectrum (a) in each figure is the inclusive spectrum of the \pik~reaction.
The middle and bottom spectra, (b) and (c), are the coincidence spectra with protons
($E_{p} > 40$~MeV) and charged pions ($E_{\pi} > 12.5$~MeV), respectively.

In Fig.~\ref{massfig_c}-(a) and \ref{massfig_si}-(a), the two dominant peaks,
denoted as s$_{\Lambda}$ and p$_{\Lambda}$ at
B$_{\Lambda} = 10.8$ and $-0.1$ MeV for \Clam, and
B$_{\Lambda} = 16.6$ and 7.1 MeV for \Silam, were interpreted
as the neutron-hole $\Lambda$-particle configurations of
$[0p^{-1}_{3/2}, s_{\Lambda}]$ and $[0p^{-1}_{3/2}, p_{\Lambda}]$ for \Clam,
and those of $[0d^{-1}_{5/2}, s_{\Lambda}]$ and $[0d^{-1}_{5/2}, p_{\Lambda}]$
for \Silam, respectively.
The quasi-free $\Lambda$ production process rises from its threshold 
(B$_{\Lambda} = 0$ MeV).
Compared to the previous hypernuclear weak decay experiments carried out at
BNL~\cite{Szym91} and KEK~\cite{Noumi95}, the background level in
Fig.~\ref{massfig_c}-(a) is greatly improved, and each peak can be clearly identified.

The peak assignments and interpretations of \Clam~and \Silam~hypernuclei
are taken from the previous spectroscopic study of light-to-heavy $\Lambda$
hypernuclei with SKS~\cite{Hase96} and theoretical investigations~\cite{Motoba88_PRC}.
Since the hypernuclear spectra of \Clam~with a resolution of 2 MeV in (FWHM)
revealed that the sub-peaks between the predominant ones have about 10\% strength
to the dominant ones, these sub-peaks were taken into account in the present peak fitting.

The fitting functions applied for the inclusive and coincidence spectra
on \Cpik~and \Sipik~reactions are as follows,
\begin{eqnarray}
  f(x) &=& \sum_{i=1}^{n}g_{i}(x - x_{i}) + QF(x) + const.
  \quad \mbox{(n=4,5 for $^{12}_{\Lambda}$C and $^{28}_{\Lambda}$Si)} \\
  g_{i}(x) &=& \frac{A_{i}}{\sqrt{2\pi}\sigma_{i}} \exp \left( -\frac{x^{2}}{2\sigma_{i}^{2}} \right) 
  \quad \left(\mbox{for the i-th peak}, \quad \sigma_{i} = \sqrt{\sigma^{2} + (\sigma_{i}^{old})^{2}} \right) \\
  QF(x) &=& \frac{1}{\sqrt{2\pi}\sigma_{1}}\int R(x - x_{th} + q)
           \exp \left( -\frac{q^{2}}{2\sigma_{1}^{2}} \right) dq \\
  R(t) &=& \left\{ \begin{array}{ll}
                          a\sqrt{t} + bt + ct^{2} + dt^{3} & \mbox{for $t>0$} \label{equ:qfree}\\
                          0                                & \mbox{otherwise}
                   \end{array} 
           \right. ,
\end{eqnarray}
where the peak position ($x_{i}$), relative intensity to the ground state ($A_{i} / A_{1}$), and
peak width ($\sigma_{i}^{old}$) of each Gaussian-like function were taken from the previous experimental
results~\cite{Hase96}.
The parameter $\sigma$ was adjusted at first by fitting the ground state 
for \Clam~and \Silam~in the proton coincidence spectra, 
representing the effect on the energy resolution since the present targets are much thicker than
those in the previous singles experiment.
The function for the quasi-free continuum ($R(t)$) was assumed to be a sum of the squared-root type function
and third-order polynomial function convoluted with the width of the ground state ($\sigma_{1}$)
as shown in Eq.~\ref{equ:qfree}.
A constant background was also assumed.
The measured energy resolutions of the ground state in the \Cpik~and \Sipik~spectra were 4.8 MeV and
6.3 MeV in FWHM, respectively.
In the fitting of the \Cpik~and \Sipik~inclusive spectra, the free parameters were the yield of
the prominent peak ($A_{1}$) and the shape of quasi-free continuum (a-d).

In Fig.~\ref{massfig_c}-(b) and \ref{massfig_c}-(c), the peak positions of two
sub-peaks (\#2 and \#3) are lower than the energy threshold of proton emission and
these states supposedly decay to the ground state by emitting gamma rays.
Since the lifetimes of these states are supposed to be on the order of 1 psec,
which is much shorter than those of the typical hypernuclear weak decay
($\tau \sim 200-300$ psec), the sub-peak events can be regarded as the ones
by the weak decay from the \Clam~ground state.
On the other hand, it is known from the old emulsion experiments that
the p$_{\Lambda}$ state of \Clam~emits a proton and sequentially produces \Blam~hypernucleus,
which then deexcites to the ground state and eventually decay by the weak interaction~\cite{Bohm}\cite{Julic72}.
Therefore, events in the p$_{\Lambda}$ state can be thought as those by the weak decay of \Blam.

It is confirmed by the previous experimental study~\cite{Ajimura_PRL92}\cite{Kishimoto_PRC95} that
quasi-free $\Lambda$ hyperons can make $\Lambda$ hypernuclei in the target.
Those $\Lambda$ hypernuclei eventually emit protons by the nonmesonic decay.
Therefore, the tagged protons in the quasi-free region can be regarded as the ones associated
with the nonmesonic decay of $\Lambda$ hypernuclei, not as misidentified pions.

In the fitting of Fig.~\ref{massfig_c}-(b) and ~\ref{massfig_c}-(c),
the yield of \plam~($A_{4}$) was also treated as free parameters.
For the peak \#2 and \#3, their relative intensity and position to the peak \#1
in the inclusive spectrum were kept constant.
The free parameters were the yields of the peak \#1 and \#4, and the shape of quasi-free
continuum.
In Fig.~\ref{massfig_si}-(b) and \ref{massfig_si}-(c), the \plam~states
(\#3 and \#4) located above the proton emission threshold can be regarded as
those of the weak decay of \Allam~after emitting a proton.
Therefore, the yields of the dominant peak (\#1, \#3, and \#5)
were treated as free parameters, while the relative peak intensity of each sub-peak (\#2 and \#4) to
the dominant one (\#1 and \#3) was fixed to be constant.

In the \Fepik~spectrum, events gated in the region of a $\Lambda$ hyperon bound
to the $^{55}$Fe nucleus were used in the analysis since each hypernuclear state could not be separated.
Considering the proton emission threshold of the excited states of \Felampre,
a series of $\Lambda$ hypernuclei ($^{56}_{\Lambda}$Fe, $^{55}_{\Lambda}$Mn,
and $^{55}_{\Lambda}$Fe) can be included in the gated region.
The symbol \Felam~represents those $\Lambda$ hypernuclei whose mass numbers are close to A$\sim$56.

The yields and statistical errors of \Clam, \Blam, \Silam, and \Allam~are
summarized in Table~\ref{yieldtab}.
The systematic errors due to the choice of functional shapes for the quasi-free region
were examined with square-root and polynomial functions.
changing the fitting region.
No significant difference was found on the results of the \slam~state of \Clam~and \Silam.
However, significant systematic changes of about 20\% for \Clam~and 50\% for \Silam
were found in the results on the \plam~states.
Such variations were included in the systematic errors of the $\pi^{-}$ branching ratios
of \Blam~and \Allam.

For \Felam~hypernuclei, the number of events in the gated window was used 
to estimate the yields.
The expected number of true pion events in the bound region of a $\Lambda$
particle was evaluated to be less than 12.4 at the 90\% confidence level (CL)
by fitting the quasi-free region in Fig.~\ref{massfig_fe}-(c).
Only an upper limit of the $\pi^{-}$ mesonic decay branching ratio
was obtained for \Felam.

\subsection{ Branching ratios of the mesonic and nonmesonic decay }

The branching ratios of the mesonic decay and the nonmesonic decay can be 
expressed by the experimental observables as follows,
\begin{eqnarray}
 b_{\pi^{-}} &=& \frac{\Gamma_{\pi^{-}}}{\Gamma_{tot}} \\
       &=& \frac{Y_{\pi^{-}}}{Y_{HY}}(\varepsilon_{coin}\Omega_{coin})^{-1}, \\
        b_{m}  &=& b_{\pi^{-}} + b_{\pi^{0}} \label{bm}, \\
        b_{nm} &=& 1 - b_{m} \label{bnm},
\label{bmesonic}
\end{eqnarray}
where the symbols of $Y_{HY}$ and $Y_{\pi^-}$ represent the yields of each
$\Lambda$ hypernucleus and associated decay pions, as described in the previous
section.
The symbol $\varepsilon_{coin}\Omega_{coin}$ denotes a product of the detection
efficiency and the solid angle of the decay counter system, which was
discussed in Sec~\ref{sec:exp}.
Table~\ref{mdbranch} shows the results of the $\pi^{-}$ mesonic decay branching 
ratios ($b_{\pi^{-}}$) of \Clam~and \Silam~hypernuclei.
An upper limit of the branching ratio of the $\pi^{-}$ mesonic decay on
\Felam~was evaluated at the 90\% CL.

\begin{figure}
\caption[]{ The pion energy spectrum of \Clam.
  The vertical axis is normalized to the total counts in the experimental data. }
\includegraphics[width=7.5cm]{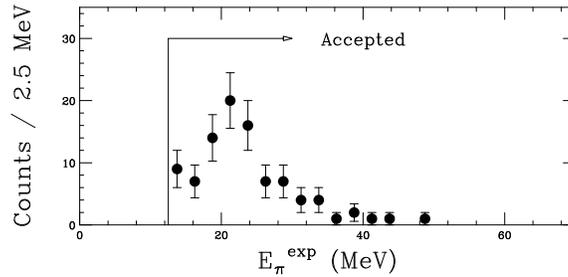}
\label{episim_c}
\end{figure}

Fig.~\ref{episim_c} shows the energy distribution of pions emitted by \Clam.
Pion energy threshold was set at 12.5 MeV.
It was found that more than 99 \% pion events in the energy spectrum were accepted
based on a simulation taking into account energy level distributions of the daughter
nucleus, $^{12}$N, calculated in ref.~\cite{Motoba88_NPA},

Branching ratios of the nonmesonic weak decay were derived incorporating those of
charged and neutral mesonic weak decay.
As for \Clam~and \Blam, the $\pi^{0}$ mesonic decay branching ratios ($b_{\pi^{0}}$) were
taken from the previous experimental data~\cite{Saka91}.
The errors of the $\pi^{0}$ branching ratios were treated as the systematic ones
in the present result.
As for \Silam~and \Allam, the ratios of the $\pi^{0}$ to $\pi^{-}$ decay width (\Gmr)
were assumed to be 2.30 for \Silam~and 0.65 for \Allam, respectively,
according to the theoretical calculations~\cite{Motoba94}.
The calculated ratios of $\Gamma_{\pi^{0}} / \Gamma_{\pi^{-}}$ were adopted to the present results.
The errors of the adopted $\pi^{0}$ branching ratios for 
\Silam~and \Allam~were assumed to be 100\%.
The mesonic decay branching ratio of \Felam~was also taken from the theoretical
calculation to be 0.015~\cite{Motoba94}.
The uncertainty of $b_{m}$ was assumed to be 100\%.

\subsection{ Proton energy spectra and \GnGp~ratios }

\begin{figure}
\caption{Comparison of the simulated energy spectra with the experimental
data.
  The horizontal axis is the observed energy by the decay counter. 
  The vertical axis is normalized by the number of hypernuclear weak decays. }
\label{epramos}
\includegraphics[width=7.5cm]{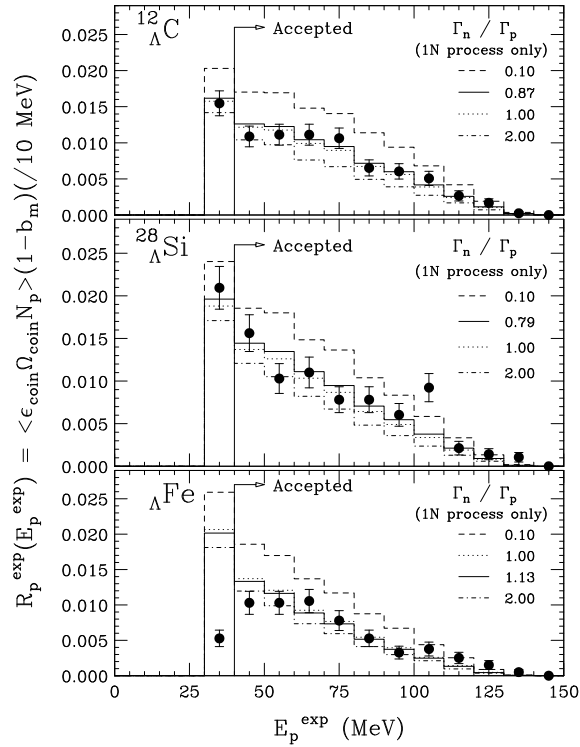}
\end{figure}

Fig.~\ref{epramos} shows the proton energy spectra measured in the present experiment.
The number of protons measured by the decay counters is plotted as a function
of the proton energy measured by the range counter ($E_{p}^{exp}$).
It is emphasized that $E_{p}^{exp}$ denotes the kinetic energy of the decay protons
at the range counter, but not the one at the weak decay vertex since
the decay protons generated at the reaction vertex undergo energy loss and struggling processes in the
experimental target and the T2 counter.
The proton energy spectrum is normalized by the number of hypernuclear weak decays
as follows,
\begin{equation}
  R_{p}^{exp}(E_{p}^{exp}) = \frac{Y_{coin}(E_{p}^{exp})}{Y_{inclusive}},
\label{Rp_Exp}
\end{equation}
where $Y_{inclusive}$ represents the number of events in the inclusive 
hypernuclear-mass spectrum, and $Y_{coin}(E_{p}^{exp})$ is the one in the 
corresponding proton-coincidence spectrum.
The error bars plotted in the spectra are statistical ones only.
The events above 40 MeV are analyzed in order to avoid the uncertainty of 
the detection limit near 30 MeV.

In accord with ref.~\cite{Ramos97}, the initial proton energy spectra at
the vertex of the nonmesonic weak decay were calculated with the pion-exchange potential
and local density approximation in finite nuclei,
assuming various \GnGp~ratios on \Clam, \Silam, and \Felam~hypernuclei.
After the nonmesonic decay, the outgoing nucleons propagate through the nucleus colliding
with other nucleons, so called the intra-nuclear cascade process (INC).
Authors of ref.~\cite{Ramos97}\cite{Ramos_pri} calculated the proton energy spectra emitted from nuclei with
help of the INC code~\cite{Carrasco}.

In ref.~\cite{Ramos97}\cite{Ramos_pri}, the proton energy spectra generated
by the proton- and neutron-induced nonmesonic decay were summed incoherently for a given \GnGp~ratio.
Explicitly, the number of protons emitted in a nonmesonic weak decay can be expressed as a function of
proton energy ($E_{p}$) and \GnGp~ratio as follows,
\begin{equation}
  N_{p}(E_{p}; \Gamma_{n} / \Gamma_{p}) = \left\{
                  \begin{array}{ll}
                         a_{n} N^{ini}_{p}(E_{p}; \Gamma_{n}) + a_{p} N^{ini}_{p}(E_{p}; \Gamma_{p})
                                    ~(a_{n} + a_{p} = 1) &
                                      \mbox{for 1N process only} \\
                         a_{n} N^{ini}_{p}(E_{p}; \Gamma_{n}) + a_{p} N^{ini}_{p}(E_{p}; \Gamma_{p})
                                                        + a_{2N} N^{ini}_{p}(E_{p}; \Gamma_{2N})\\
                                    ~(a_{n} + a_{p} + a_{2N} = 1) &
                                      \mbox{for 1N + 2N process}
                  \end{array}
                  \right. ,
\label{npramos}
\end{equation}
where $a_{n}$, $a_{p}$, and $a_{2N}$ are the fractions of neutron- and proton-induced
nonmesonic decay ($\Lambda \rightarrow N N$), and the two-nucleon induced decay
($\Lambda N N \rightarrow N N N$), respectively.
$N^{ini}_{p}(E_{p}; \Gamma_{n})$ and $N^{ini}_{p}(E_{p}; \Gamma_{p})$ are the numbers of protons
originated in the neutron- and proton-induced process, so called ``1N process'',
and $N^{ini}_{p}(E_{p}; \Gamma_{2N})$ is that originated in the two-nucleon induced decay,
so called ``2N process'', respectively.
The number of protons in the above equation is normalized
to be the one per nonmesonic weak decay of a $\Lambda$ hypernucleus.

The proton energy spectra measured by the decay counters were calculated by the GEANT simulation
code described in the previous section.
In the Monte-Carlo simulation, energetic protons emitted from the target nucleus
 with the energy distributions of protons ($N_{p}(E_{p}; \Gamma_{n} / \Gamma_{p})$) were 
transported through the target material and the decay counters.
The geometrical acceptance, the energy loss, and its straggling in the target and decay counters
together with the energy threshold of 40 MeV were automatically taken into account.
The notation of $<\Omega_{coin}N_{p}(E_{p}; \Gamma_{n}/\Gamma_{p})>$ is introduced
to represent the proton energy spectra obtained by the GEANT simulation code.

The simulated proton energy spectrum ($R_{p}^{cal}(E_{p}^{exp})$) to be compared with the
experimental data ($R_{p}^{exp}(E_{p}^{exp})$) was obtained as follows,
\begin{eqnarray}
 R_{p}^{cal}(E_{p}^{exp}) &=& 
 <\Omega_{coin}N_{p}(E_{p}; \Gamma_{n}/ \Gamma_{p})>\varepsilon_{coin}b_{nm},
\label{Epnorm}
\end{eqnarray}
where $\varepsilon_{coin}$, $\Omega_{coin}$, and $b_{nm}$ represent 
the detection efficiency, the solid-angle acceptance of the decay counters,
and the branching ratio of the nonmesonic decay described in the previous
section, respectively.

The \GnGp~ratios of \Clam, \Silam, and \Felam~were obtained by fitting the 
experimental data with the simulated spectra, changing the \GnGp~ratio
as a free parameter.
The results in both cases of the ``1N only'' and ``1N and 2N'' processes
are listed in Table~\ref{nmdrate}.
The simulated proton energy spectra in the case of 
\GnGp = 0.1, 1.0, and 2.0 are overlaid on the experimental data
in Fig.~\ref{epramos}. 
It is shown that the proton yield is much smaller than the calculated one
assuming \GnGp = 0.1, which seems to exclude very small values of \GnGp.

Recently, an erratum of ref.~\cite{Ramos97} was published,
giving the new proton energy spectra~\cite{Ramos97_erratum}.
Since much stronger final state interaction in the INC process
was applied in the updated spectra than that in the original paper,
the shape of the spectra was greatly shifted toward the lower energy side.
The present results of the \GnGp~ratios listed in Table~\ref{nmdrate}
are updated from the ones published in our previous report~\cite{E307pdw_letter}
by applying the new proton energy spectra given by the authors of
ref.~\cite{Ramos97_erratum} to the experimental data.
It should be noted that the experimental proton energy spectra published
in ref.~\cite{E307pdw_letter} themselves do not change at all
in the present report.
Since the experimental data of proton energy spectra noted in Eq.~(\ref{Rp_Exp})
contains only experimental observables, they are free from theoretical assumptions.
As a result, the \GnGp~ratios of \Clam~and \Silam~became smaller than
our previous ones published in ref.~\cite{E307pdw_letter}.

The statistical error of each \GnGp~ratio for \Clam, \Silam, and \Felam~was
evaluated by the conventional least-square method.
In addition, the systematic error was evaluated, considering the uncertainty
of the nonmesonic decay branching ratio, the detection efficiency of the decay
counters, and the solid angle acceptance. 
The systematic errors of the ratio $R_{p}^{cal}(E_{p}^{exp})$ in the equation (\ref{Epnorm})
were found to be 0.09, 0.10, and 0.05 for \Clam, \Silam, and \Felam, respectively. 
For \Felam, the dependence of the gate position for the $\Lambda$
bound state on the \GnGp~ratio was also examined.
The differences of the obtained \GnGp~ratios for 160 MeV $< M_{HY}-M_{A} <$ 180 MeV
and 160 MeV $< M_{HY}-M_{A} <$ 190 MeV were found to be 13 \% for the 1N process
and 15 \% for the 1N+2N process. 
These errors were included in the systematic errors for \Felam.

Assuming that the decay widths of the 1N and 2N process are 0.803 and
0.277 on \Clam~\cite{Ramos97}\cite{Ramos_pri}, the \GnGp~ratio~on \Clam~is
$0.60^{+0.11 +0.23}_{-0.09 -0.21}$.
Recently, Alberico {\it et al.} evaluated the decay width of the 2N process
to be 0.16 on \Clam, based on the propagator method
\cite{Alberico_PRC00}\cite{Alberico_HYP2000}.
When we take this value, the \GnGp~ratio of \Clam~changes from 0.60 to 0.70.
Since most of protons emitted by the 2N process are not detected by our decay
counter systems, the inclusion of the 2N process in the fitting gives smaller
\GnGp~ratios than those in the case of 1N process only.
The systematic error of the present data listed in Table~\ref{nmdrate} was
evaluated only by the experimental conditions described above.
The uncertainties of the derived \GnGp~ratios due to theoretical assumptions and
models, such as degree of the 2N process contribution and the ambiguity coming
from the intranuclear cascade calculations, are not included in the systematic
errors quoted in the present results.

\section{ Discussion }
\label{sec:discuss}

\subsection{ Mesonic decay widths }

\begin{figure}
\caption[]{ Comparison of the existing and present data with theoretical 
calculations by Motoba {\it et al.}~\cite{Motoba88_NPA}\cite{Motoba94} and
Nieves {\it et al.}~\cite{Nieves93}.
``FREE'' uses a plane wave and only the Coulomb potential for the pion wave 
function. ``MSU'' uses an optical potential developed by a group of Michigan State
University~\cite{MSU}. ``WHIS'' uses an optical potential developed by Whisnant
\cite{WHIS}. ``Nieves'' uses a theoretical calculation by Nieves {\it et al.}. 
``Szym'' and ``Noumi'' are the previous experimental data by
 Szymanski {\it et al.}~\cite{Szym91} and Noumi {\it et al.}~\cite{Noumi95},
 respectively.
 The closed circles are the present results.}
\label{newgamma_pi2}
\includegraphics[width=7.5cm]{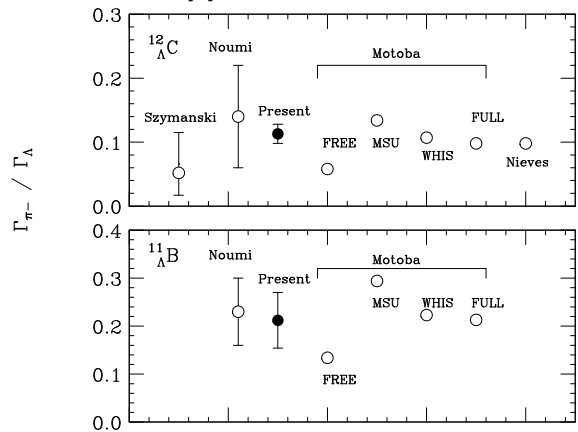}
\end{figure}

Table~\ref{mdrate} and Fig.~\ref{newgamma_pi2} summarize the present results
for the mesonic decay widths together with those obtained by
Szymanski {\it et al.}~\cite{Szym91} and Noumi {\it et al.}~\cite{Noumi95}.
The results of \Gpim~were derived from the obtained branching ratios using the 
total decay widths measured simultaneously in the present experiment
\cite{Bhang98}\cite{Park99} by the relation, 
$\Gamma_{\pi^{-}} = \Gamma_{tot} b_{\pi^{-}}$.
The present $\pi^{-}$ decay widths of \Clam~and \Blam~hypernuclei are 
much more precise than the previous experimental data, 
and allow us a critical comparison with theoretical calculations.
Furthermore, finite values of \Gpim~for \Silam~and \Allam~were measured
for the first time.
For \Felam, an upper limit of the $\pi^{-}$ mesonic decay width was obtained.

In ref.~\cite{Motoba88_NPA}, Motoba {\it et al.} calculated the mesonic decay
widths of \Clam~and \Blam~with the pion-nucleus optical potentials given by a group of
Michigan State University (MSU)~\cite{MSU} and Whisnant~\cite{WHIS},
which have a strong and weak imaginary part, respectively.
Another potential was introduced to resolve the singularity in the pion wave
function (FULL)~\cite{Motoba94}.
The calculated results are compared to the calculations with the plane wave 
including the Coulomb distortion (FREE).
In addition, Nieves {\it et al.} has developed an optical potential,
which was derived from a purely theoretical calculation, and can reproduce
the experimental data of pionic atoms and $\pi$-nucleus scattering
data~\cite{Nieves93}.
They applied their optical potential to the mesonic decay of $\Lambda$ 
hypernuclei incorporating the nuclear shell model wave functions.
A similar comparison of the experimental results for the $\pi^{-}$ mesonic 
decay of \Silam~and \Allam~with the theoretical results is also
shown in Table~\ref{mdrate} only for the case of ``FULL''.

It can be said that theoretical calculations taking into account of
the distortion effects of the pion wave function and nuclear wave functions
agree well with the present experimental results.
In particular, the result on \Clam, whose experimental error has been drastically improved, 
is consistent with an increase of \Gpim~due to the distortion of the pion
wave function by the pion-nucleus optical potentials.

\subsection{ Nonmesonic decay widths and \GnGp~ratios}

\begin{figure}
\caption[]{ Total nonmesonic decay widths of $\Lambda$ hypernuclei.
The open circles are previous experimental data in which the hypernuclear production
was explicitly identified.
The open diamonds are experimental data by the p+Bi and p+U reactions 
with the recoil shadow method in which the production of strangeness was not
explicitly identified.
The close circles are the present results. 
The solid line is a calculation by Itonaga {\it et al.}~\cite{Itonaga_PRC02}.
The dash-dotted line is a calculation by Ramos {\it et al.}~\cite{Ramos94}.
The dashed line is a calculation by Alberico {\it et al.}~\cite{Alberico_PRC00}
($\Gamma_{1N}$ only).
The open square shows a result with the direct quark exchange model
in nuclear medium by Sasaki{\it et al.}~\cite{Oka_NPA00}.}
\label{nmdecay_adep}
\includegraphics[width=7.5cm]{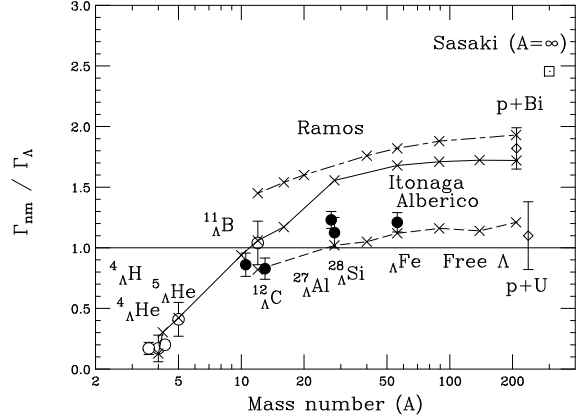}
\end{figure}

Table~\ref{nmdrate} summarizes experimental results of nonmesonic decay width
and \GnGp~ratios along with theoretical ones.
Fig.~\ref{nmdecay_adep} shows the mass-number dependence of the nonmesonic
weak decay widths derived by the present and previous experiments.
There are experimental data of lifetime measurement for very heavy hypernuclei
in the mass region of $A \sim 200$ with recoil shadow method on p+Bi and p+U reactions
at COSY, although the strangeness production was not explicitly identified
\cite{COSY_PRC97}\cite{COSY_NPA98}.
The plotted data around $A \sim 200$ in Fig.~\ref{nmdecay_adep} were the ones
converted from the results of lifetime measurements, assuming that no mesonic decay
occurs in heavy $\Lambda$ hypernuclei.
Theoretical calculations of the nonmesonic decay widths by Itonaga {\it et al.}
\cite{Itonaga_PRC02}, Ramos {\it et al.}~\cite{Ramos94},
Alberico {\it et al.}~\cite{Alberico_PRC00}, and Sasaki {\it et al.}~\cite{Oka_NPA00}
are also illustrated in the figure.

The calculation by Ramos {\it et al.}~\cite{Ramos94}, on which
present derivation of \GnGp~ratios rely,  was based on one-pion exchange potential
in which the vertex renormalization effect in the nuclear medium and
the local density approximation are taken into account. 
Their results of the nonmesonic decay widths are much larger than the
experimental results, and do not seem to saturate around A=56.
Recently, Alberico {\it et al.}~\cite{Alberico_PRC00} updated Ramos's 
calculation, and obtained results closer to the present
data by adjusting the Landau-Migdal parameter, which
controls the short-range part of the pion potential.

\begin{figure}
\caption[]{ The \GnGp~ratios of $\Lambda$ hypernuclei.
The closed and open circles are the results of the present experiment,
assuming the ``1N only'' and ``1N and 2N'' processes, respectively.
The open squares are previous experimental data by
Szymanski {\it et al.}~\cite{Szym91} and Noumi {\it et al.}~\cite{Noumi95}. 
Theoretical calculations by Itonaga~\cite{Itonaga_PRC02},
Parreno~\cite{Parreno_PRC01}, and Sasaki~\cite{Oka_NPA00} are plotted.}
\label{gngp_all}
\includegraphics[width=7.5cm]{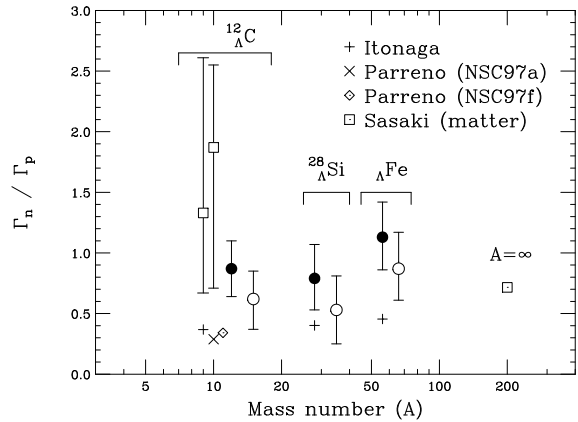}
\end{figure}

The present data substantially improved the knowledge of the \GnGp~ratios for
\Clam~and \Blam~as shown in Fig.~\ref{gngp_all}.
Although all available experimental data agree with the present ones within 1 sigma level,
the present \GnGp's are in the region around 0.5-1.0, while the previous ones
were significantly larger than unity.

The direct quark current exchange model (DQ) was applied
to explain the very short-range part of the nonmesonic decay
in light $\Lambda$ hypernuclei and nuclear matter~\cite{Oka}\cite{Oka_NPA00},
incorporating it with the contribution of $\pi$ and $K$ meson exchanges.
Parreno {\it et al.} carried out a systematic study about various combinations
of meson exchange potentials, such as $\pi, K, \eta, \rho$, and $\omega$
mesons~\cite{Parreno_PRC01}, which contains an update of the results published
in ref.~\cite{Parreno_PRC97}.
Itonaga {\it et al.} introduced additional potentials,
V$_{2\pi/\rho}$ and V$_{2\pi/\sigma}$, in which terms of correlated
$2\pi/\rho$ and $2\pi/\sigma$ exchanges are taken into account~\cite{Itonaga_PRC02}
since it has a strong tensor force but still has opposite sign to that of the one-pion
exchange potential.
It was also pointed out that two-nucleon induced nonmesonic decay ($\Lambda N N \rightarrow N N N$)
could play an important role in the nonmesonic weak decay~\cite{Alberico_PRC00}.
However, no experiment has so far explicitly measured the contribution of the two nucleon process.

The present results of the nonmesonic decay widths and the \GnGp~ratios
provide good criteria to test the short-range nature of the nonmesonic decay,
which is essentially important to understand its mechanism.
However, as shown in Fig.~\ref{nmdecay_adep} and \ref{gngp_all}, there is no
theoretical calculation which explains both the nonmesonic decay
width and the \GnGp~ratio consistently.
For instance, the quark-current exchange model gives the \GnGp~ratio
in nuclear matter comparable to the present experimental data, but overestimates the
nonmesonic decay width.
The conventional meson exchange models, in which the contribution of heavy mesons
are included, provide larger \GnGp~ratios than those by the one-pion exchange model,
though the results of meson exchange models are still smaller than the experimental data.

From an experimental point of view, more exclusive information, such as
the angular correlation of outgoing nucleons and the detection of low-energy
neutrons, would also help to solve the puzzle of the \GnGp~ratio.
Recently, an experiment to measure the neutron spectra of the nonmesonic weak decay
of \Clam~and \Ylam~was carried out at KEK~\cite{E369exp}\cite{E369_Hotchi}.
The observed number of neutrons in \Clam~\cite{E369_JHKim} was reported to be
smaller than that expected from ref.~\cite{Ramos97}\cite{Ramos97_erratum}.
In addition, coincidence measurements of the decay particles including
neutrons from the s- and p-shell hypernuclei, \Hegolam~and \Clam, were
carried out at KEK~\cite{E462exp}\cite{E462_Bhang}, and the data analysis is
under way.
A study of the correlation of pair nucleons emitted in the nonmesonic weak decay
with much greater statistics would help to completely understand the nucleon
spectra and to reduce any ambiguity in the final state interaction.

\section{ Conclusion }
\label{sec:concl}

In the present study, we measured the pion and proton energy spectra,
and extracted the $\pi^{-}$ mesonic decay widths (\Gpim), the total nonmesonic decay
widths (\Gnm), and the \GnGp~ratios on medium-to-heavy $\Lambda$ hypernuclei
with high precision.
The results of \Gpim~on \Clam~confirmed an increase of
the mesonic decay width, which is interpreted as due to the distorted pion
wave function and the nuclear shell-model configurations.
The total nonmesonic decay widths (\Gnm) in medium-to-heavy $\Lambda$
hypernuclei almost saturates at $A \sim 56$.
This suggests that the short-range nature and the local density approximation
take central roles in the mechanism of the nonmesonic decay.
The present \GnGp~ratios are now comparable with the recent calculations based on the
meson exchange potentials taking into account of heavy mesons such as Kaon, $\rho$,
and $\omega$, and/or the quark-current.
Although the experimental uncertainty in deriving the \GnGp~ratios become less than
ever, the experimental \GnGp~ratios are still larger than the theoretical
calculations.
At this moment, there is no theoretical calculation that can reproduce
both the nonmesonic decay widths and the \GnGp~ratios.
Further experimental and theoretical efforts are needed to fully reveal
the mechanism of the nonmesonic weak decay.

\acknowledgments

The authors thank Professor K.~Nakai and Professor T.~Yamazaki for their
continuous encouragement throughout the experiment from the beginning.
We deeply appreciate the staff members of KEK and INS for their generous support
to carry out the experiment and data analysis.
Discussions with Professor K.~Itonaga, Professor T.~Motoba,
Professor T.~Fukuda, and Professor T.~Nagae are greatly appreciated.
One of the authors (H.B.) acknowledges partial support from the Korea-Japan
collaborative research program of KOSEF (R01-2000-000-00019-0) and
the Korea Research Foundation (2000-015-Dp0084).

%
%

\newpage

\newpage

\newpage

\newpage

\newpage

\newpage

%
%
\clearpage
\narrowtext

\begin{table}[htb]
\caption[]{ The number of incident pions used in the present analysis and
the estimated yields of each hypernuclear state.}
\begin{center}
\begin{tabular}{rcccc}
\hline\hline
      & $N_{\pi^{+}}$ ($\times 10^{9}$) & Y$_{HY}$ & Y$_{p}$ & Y$_{\pi^{-}}$\\
\hline
\Clam  & 314     & $4132 \pm 70$ & $273 \pm 17$ & $93 \pm 10$ \\
\Blam  &         & $4016 \pm 92$ & $213 \pm 20 \pm 18$ & $155 \pm 24 \pm 32$\\
\Silam & 620     & $2631 \pm 74$ & $190 \pm 15$        & $23 \pm 5$ \\
\Allam &         & $3693 \pm 83$ & $218 \pm 20$        & $28 \pm 7 \pm 13$ \\
\Felam & 919     & $3981 \pm 96$ & $222 \pm 15$ & $23 \pm 5$ \\
\hline\hline
\end{tabular}
\end{center}
\label{yieldtab}
\end{table}

\begin{table}[htb]
\caption{ The $\pi^{-}$ mesonic decay branching ratios ($b_{\pi^{-}}$) obtained
in the present analysis and the mesonic decay branching ratios 
($b_{m}=b_{\pi^{-}} + b_{\pi^{0}}$).
The quoted errors are statistical and systematic, respectively.}
\label{mdbranch}
\begin{tabular}{cccc}
\hline\hline
              & b$_{\pi^{-}}$      & b$_{\pi^{0}}$ & b$_{m}$ \\
              & ($\times 10^{-2}$) & ($\times 10^{-2}$) & ($\times 10^{-2}$)\\
\hline
\Clam         & $ 9.9 \pm 1.1 \pm 0.4$ & $17.4 \pm 5.7 \pm 0.8$~\cite{Saka91} 
              & $27.3 \pm 1.1 \pm 5.8$ \\
\Blam         & $17.0 \pm 2.7 \pm 3.6$ & $14.0 \pm 3.9 \pm 2.5$~\cite{Saka91} 
              & $31.0 \pm 2.7 \pm 5.9$ \\
\Silam        & $3.6  \pm 0.8 \pm 0.2$ & $ 8.3 \pm 8.3^{a)}$              
              & $11.9 \pm 0.8 \pm 8.3^{a)}$ \\
\Allam        & $3.2  \pm 0.8 \pm 1.5$  & $ 2.0 \pm 2.0^{a)}$              
              & $5.2  \pm 0.8 \pm 2.5^{a)}$ \\
 
\Felam        & $<1.2$ (90\% CL)   &                & $1.5 \pm 1.5$~\cite{Motoba94} \\
\hline\hline
\end{tabular}\\
a) The ratios of $\Gamma_{\pi^{0}} / \Gamma_{\pi^{-}}$ were assumed 
to be 2.30 for \Silam~and 0.65 for \Allam~with 100\% errors, according 
to the theoretical calculations by Motoba {\it et al.}~\cite{Motoba94}.
\end{table}

\begin{table}[htb]
\caption{ Nonmesonic decay widths and \GnGp~ratios in the present 
experiment. All the decay widths are listed in units of the total decay width of
a $\Lambda$ hyperon in free space (\Glam).}
\label{nmdrate}
\begin{tabular} {ccccc}
\hline\hline
        & \Gnm / \Glam & \multicolumn{2}{c}{\GnGp} & Refs.\\
        &              & ``1N only'' & ``1N and 2N'' & \\
\hline
Experiment & & & &\\
\Clam  & $0.828 \pm 0.056 \pm 0.066$ & 
$0.87 \pm 0.09 \pm 0.21$ & 
$0.60^{+0.11 +0.23}_{-0.09 -0.21}$ & Present$^{a)}$ \\
& & & ($\Gamma_{1N} = 0.80, \Gamma_{2N} = 0.28$)~\cite{Ramos97} &\\
       & $0.89 \pm 0.15 \pm 0.03$ & $1.87 \pm 0.59^{+0.32}_{-1.00}$ & 
& \cite{Noumi95} \\
       & $1.14 \pm 0.20$          & $1.33^{+1.12}_{-0.81}$ & & \cite{Szym91} \\
\Blam  & $0.861 \pm 0.063 \pm 0.073$ & & & Present \\
       & $0.95 \pm 0.13 \pm 0.04$ & $2.16 \pm 0.58^{-0.45}_{-0.95}$ & 
& \cite{Noumi95} \\
       &                          & $1.04^{+0.59}_{-0.48}$ & & \cite{Szym91} \\
\Silam & $1.125 \pm 0.067 \pm 0.106$ & 
$0.79^{+0.13 +0.25}_{-0.11 -0.24}$ & 
$0.53^{+0.13 +0.25}_{-0.12 -0.24}$ & Present$^{a)}$ \\
& & & ($\Gamma_{1N} = 0.94, \Gamma_{2N} = 0.28$)~\cite{Ramos97} & \\
\Allam & $1.230 \pm 0.062 \pm 0.032$ & & & Present \\
\Felam & $1.21 \pm 0.08$           &
$1.13^{+0.18 +0.23}_{-0.15 -0.22}$ & 
$0.87^{+0.18 +0.23}_{-0.15 -0.21}$ & Present$^{a)}$ \\
 & & & ($\Gamma_{1N} = 1.02, \Gamma_{2N} = 0.27$)~\cite{Ramos97} &\\
Theory & & & \\
\Clam       & 1.060 & 0.368 & & \cite{Itonaga_PRC02} \\
            & 0.82  &       & & \cite{Alberico_PRC00} \\
            & 0.726 & 0.288 & & NSC97a~\cite{Parreno_PRC01} \\
            & 0.554 & 0.341 & & NSC97f~\cite{Parreno_PRC01} \\
\Silam      & 1.556 & 0.402 & & \cite{Itonaga_PRC02} \\
            & 1.02  &       & & \cite{Alberico_PRC00} \\
\Felampre   & 1.679 & 0.455 & & \cite{Itonaga_PRC02} \\
            & 1.12  &       & & \cite{Alberico_PRC00} \\
$p + ^{209}$Bi & $1.82 \pm 0.09 \pm 0.29^{b)}$ & & & \cite{COSY_NPA98}\\
$p + ^{238}$U  & $1.10 \pm 0.28^{b)}$ & & & \cite{COSY_PRC97} \\
$A=\infty$  & 2.456 & 0.716 & & $\pi + K + DQ$~\cite{Oka_NPA00} \\
\hline\hline
\end{tabular}
a) The uncertainty of theoretical calculations in the proton energy spectra
is not included in the systematic errors.\\
b) The results of lifetime measurements were converted to the nonmesonic
decay widths, assuming that no mesonic decay occurs in heavy $\Lambda$ hypernuclei.
\end{table}

\begin{table}
\caption{ Mesonic decay widths with theoretical calculations.
All the decay widths are listed in units of the total decay width of
a $\Lambda$ hyperon in free space (\Glam).}
\label{mdrate}
\begin{tabular} {ccc}
\hline\hline
       & \Gpim / \Glam               &  Refs. \\
\hline
Experiment & & \\
\Clam  & $0.113 \pm 0.014 \pm 0.005$ & Present \\
       & $0.14  \pm 0.07  \pm 0.03 $ & \cite{Noumi95} \\
       & $0.052^{+0.063}_{-0.035}$   & \cite{Szym91} \\
\Blam  & $0.212 \pm 0.036 \pm 0.045$ & Present \\
       & $0.23  \pm 0.06  \pm 0.03 $ & \cite{Noumi95} \\
\Silam & $0.046 \pm 0.011 \pm 0.002$ & Present \\
\Allam & $0.041 \pm 0.010 \pm 0.019$ & Present \\
\Felam & $< 0.015$ (90\% CL)         & Present \\
Theory & & \\
\Clam  & 0.058 (FREE) & \cite{Motoba88_NPA} \\
       & 0.134 (MSU)  & \cite{Motoba88_NPA} \\
       & 0.107 (WHIS) & \cite{Motoba88_NPA} \\
       & 0.098 (FULL) & \cite{Motoba94} \\
       & 0.086 (Nieves) & \cite{Nieves93} \\
\Blam  & 0.134 (FREE) & \cite{Motoba88_NPA} \\
       & 0.294 (MSU)  & \cite{Motoba88_NPA} \\
       & 0.223 (WHIS) & \cite{Motoba88_NPA} \\
       & 0.213 (FULL) & \cite{Motoba94} \\
\Silam & 0.027 (FULL) & \cite{Motoba94} \\
\Allam & 0.065 (FULL) & \cite{Motoba94} \\
\hline\hline
\end{tabular}
\end{table}

\end{document}